\documentclass{emulateapj}

\usepackage{txfonts} 
\usepackage{natbib} 
\usepackage{graphicx}
\usepackage{epstopdf}

\newcommand{\reaction}[6]{\nuc{#1}{#2}(#3,#4)\/\nuc{#5}{#6}}
\newcommand{\nuc}[2]{\ensuremath{^{#1}}#2}

\begin{document}

\title{Nucleosynthesis during the Merger of White Dwarfs and the
  Origin of R Coronae Borealis Stars}

\author{R.~Longland$^{1,2}$, P.~Lor\'{e}n-Aguilar$^{3,2}$,
J.~Jos\'{e}$^{1,2}$, E. Garc\'{\i}a--Berro$^{3,2}$,
L.~G.~Althaus$^{4}$, \& J.~Isern$^{5,2}$}
        
\affil{$^1$Departament de F\'{\i}sica i Enginyeria Nuclear, EUETIB,
  Universitat Polit\`{e}cnica de Catalunya, c/ Comte d'Urgell 187,
  E-08036 Barcelona, Spain\\
  $^2$Institut d'Estudis Espacials de Catalunya (IEEC), Ed. Nexus-201,
  C/ Gran Capit\`{a} 2-4, E-08034
  Barcelona, Spain\\
  $^3$Departament de F\'{\i}sica Aplicada, Universitat Polit\`{e}cnica
  de Catalunya, c/ Esteve Terrades, 5,
  E-08860 Castelldefels, Spain\\
  $^4$Facultad de Ciencias Astron\'omicas y Geof\'{\i}sicas,
  Universidad Nacional de La Plata, Paseo del Bosque
  s/n, (1900) La Plata, Argentina\\
  $^5$Institut de Ci\`encies de l'Espai (CSIC), Campus UAB, E-08193
  Bellaterra, Spain}

\email{richard.longland@upc.edu}

\begin{abstract} 
  Many hydrogen deficient stars are characterised by surface abundance patterns that are hard to reconcile with conventional stellar evolution. Instead, it has been suggested that they may represent the result of a merger episode between a helium and a carbon-oxygen white dwarf.  In this Letter, we present a nucleosynthesis study of the merger of a $0.4\, M_{\sun}$ helium white dwarf with a $0.8\, M_{\sun}$ carbon-oxygen white dwarf, by coupling the thermodynamic history of Smoothed Particle Hydrodynamics particles with a post-processing code.  The resulting chemical abundance pattern, particularly for oxygen and fluorine, is in qualitative agreement with the observed abundances in R Coronae Borealis stars.
\end{abstract}


\maketitle

\section{Introduction}

Hydrogen deficient stars with high carbon abundances (with the
exceptions of the PG 1159 and CSPN WR stars) can be divided into three
main categories according to their effective temperatures: Hydrogen
Deficient Carbon (HdC) stars, R~Coronae~Borealis (RCB), and Extreme
Helium (EHe) stars.  They are often considered to be at different
evolutionary stages following a common origin, as evidenced by their
similar atmospheric abundances \citep{PAN04}.

The origin of these stars is of particular interest because of their
unique chemical composition.  As well as being enriched in carbon and
oxygen relative to solar abundances, they also exhibit enrichment in
{nitrogen, neon, fluorine, and some s-processed
  material}. These abundance patterns are hard to reconcile with their
expected initial composition, requiring alternative scenarios for
their origin. Two leading theories are the Final Flash (FF) and the
Double-Degenerate (DD) scenarios.  The FF scenario attributes the
formation to a late He-shell flash in a post-Asymptotic Giant Branch
(post-AGB) star as it evolves towards {the white dwarf
  state}. The flash would move the star back onto the AGB part of the
Hertzsprung Russell diagram, while mixing s-processed material and
He-burning ashes to the surface \citep{FUJ77,REN79,WER06}.  In the DD
scenario, the merger of two white dwarfs (He+CO), with a total mass
below the Chandrasekhar limit, through disruption and accretion of the
lighter star, gives rise to a single object whose composition is an
admixture of the two white dwarfs \citep{WEB84,ASP00}.  This dense
object is expected to be surrounded by a hot corona and an accretion
disk \citep{LOR09}. Surface abundance analyses of hydrogen-deficient
stars seem to favour the latter scenario\ \citep{SAI02}, although the
possibility of some objects being formed through the FF scenario
cannot be excluded \citep{PAN11}.

When considering the formation of hydrogen deficient objects through
the DD scenario, the following question can be raised: does nuclear
processing occur during the merging event?  If no nuclear burning
occurs --- the ``cold'' merger hypothesis --- the observed abundances
might be reproduced through the {partial} mixing of material previously
processed during the AGB phase of the CO white dwarf
progenitor. However, recent calculations based on this scenario failed
to qualitatively match the overall abundance pattern observed
\citep{JEF11}.  Another possibility involves a ``hot'' merger event,
in which significant nuclear processing occurs.  Indeed, \cite{CLA07}
suggested that this scenario may account for the high abundances of
$^{18}$O observed in some stars.  Unfortunately, no detailed
nucleosynthesis study during white dwarf mergers has been published to
date.

In this Letter, we present a detailed study of the nucleosynthesis
accompanying white dwarf mergers. Calculations are based on
post-processing analyses using temperature and density versus time
profiles obtained from Smoothed Particle Hydrodynamics simulations of
the merger event. The structure of the manuscript is as follows: in
section \ref{sec:observations} we summarise the observed abundances in
hydrogen deficient objects. The details of previous nucleosynthesis
studies and of the calculations presented in this work are given in
sections \ref{sec:previous-studies} and \ref{sec:SPH}, respectively.
Results are discussed in section \ref{sec:nucleosynthesis}, and the
main conclusions are reached in section \ref{conclusions}.

\section{Abundances in H Deficient Stars}
\label{sec:observations}

{All current observed surface abundances of hydrogen deficient stars
have been collected recently by \cite{JEF11}\footnote{with the
  exception of $^{12}$C$/^{13}$C and $^{16}$O$/^{18}$O ratios, and
  abundances of Sc, Mn, Co, Cu, and Sr}}, to which the reader is
referred for details.  Here we summarise the main observational
features reported in that work.

Carbon is enriched in all observed hydrogen deficient stars,
irrespective of their iron content, with overabundances relative to
solar of the order of 0.7 dex.  Additionally, the inferred
$^{12}$C$/^{13}$C ratios are very large ($> 500$), indicating helium
burning in the star. Nitrogen is also enriched in most EHe and RCB
stars. However, in this case it is correlated to the iron abundance,
suggesting that nitrogen is enriched only by CNO cycling prior to the
merging event, as can be shown by summing the initial carbon, oxygen,
and nitrogen abundances of the star.  Oxygen poses an interesting
problem since its abundance exhibits a large spread.  Regardless of
this fact, oxygen appears to be both enriched and independent of
metallicity.  To account for the overabundance of oxygen a dredge-up
event, by which oxygen is injected from the CO white dwarf, seems to
be required.  Because of the difficulties in distinguishing between
isotopes in hotter stars, oxygen could also be produced in the form of
$^{18}$O, which has been found to be highly enriched in a few stars
\citep{CLA07}.  Fluorine, in turn, is highly enriched by 2--4 dex in
all the observed stars. This enrichment could possibly arise from
processing during the AGB phase of the CO white dwarf progenitor
through a rather complicated process in the helium inter-shell zone
\citep{LUG04}. Finally, s-process elements, particularly yttrium and
zirconium, have also been observed to be enriched by up to a factor of
50 in some EHe stars \cite{PAN04}. The source of these elements is
probably dredge-up of helium inter-shell products of the AGB star that
subsequently evolved into the CO white dwarf. While several models can
match some of these observational features to an order of magnitude
level, the current challenge is to reproduce the overall abundance
pattern simultaneously with a detailed model of the merger event.

\section{Previous Nucleosynthesis Studies}
\label{sec:previous-studies}

Most abundance studies of white dwarf mergers performed so far
\citep{SAI02,PAN11,JEF11} {consider} the cold merger paradigm, in which
no nucleosynthesis occurs.  In particular, the recent work by
\cite{JEF11} started with the abundances present in the white dwarfs
as resulting from previous evolution, particularly the AGB star
evolution of the CO white dwarf progenitor. Calculations also included
helium and hydrogen buffers on the surfaces of the white dwarfs,
remaining from previous evolution, which contained material that had
been heavily processed.  By using a recipe for mixing the material in
the white dwarfs during the merger, they were able to account for most
of the abundances (or at least the trends) observed in EHe and RCB
stars. The most notable exception, however, was the abundance of
fluorine. Although fluorine can exist in the helium buffer of the CO
white dwarf, it was found not to be abundant enough to account for the
very high observed values.

Hot merger nucleosynthesis calculations were performed by \cite{CLA07}
upon the discovery of very low $^{16}$O$/^{18}$O ratios in some HdC
and RCB stars. These studies relied on a one-zone parametric model to
simulate the conditions needed to burn helium during a merger
event. Their simple model was able to account for the enrichment of
$^{18}$O in these environments. When an admixture of hydrogen was
added, however, the $^{18}$O was rapidly destroyed through proton
captures while $^{16}$O was created through the
$^{13}$C$(\alpha,n)^{16}$O reaction, thus restoring the
$^{16}$O$/^{18}$O ratio closer to solar values.  Hot merger
nucleosynthesis was also studied by \cite{JEF11} in the absence of
mixing of material from the CO white dwarf. They were able to account
for overabundances in carbon and oxygen, but limitations in {the SPH
calculations provided by \cite{LOR09}} prevented them from making
reliable predictions for other nuclei. Detailed nucleosynthesis
studies using realistic hydrodynamic models are therefore required to
fully understand the production of elements in these environments.

\section{Initial Models and Input Physics}
\label{sec:SPH}

The models used as input for the nucleosynthesis calculations reported
in this paper have been presented elsewhere \citep{GUE04,LOR09,LOR10},
and rely on a Smoothed Particle Hydrodynamics (SPH) Lagrangian
simulation to follow the evolution of two coalescing white dwarfs.  We
refer the reader to these works for the details of the dynamical
behaviour of the merger process, but we describe here the structure of
the final remnant, which is relevant for the calculations reported
below.  In essence, the final remnant of the merger process consists
of a central hot white dwarf which contains most of the mass of the
primary.  On top of it a hot corona can be found, which contains most
of the mass of the disrupted secondary and a small admixture of the
primary white dwarf.  Finally, surrounding the compact remnant, a
rapidly rotating Keplerian disk is formed.  It is also worth noting
that little mass is ejected from the system.  We emphasise that, for
the sake of conciseness, we only describe the results obtained when
the masses of the merging white dwarfs are 0.4 and $0.8\, M_{\sun}$
{with SPH particle masses of $2.6 \times 10^{-6} M_{\sun}$ and $5.3
\times 10^{-6} M_{\sun}$, respectively.}

\begin{figure} \centering
  \includegraphics[width=0.7\hsize]{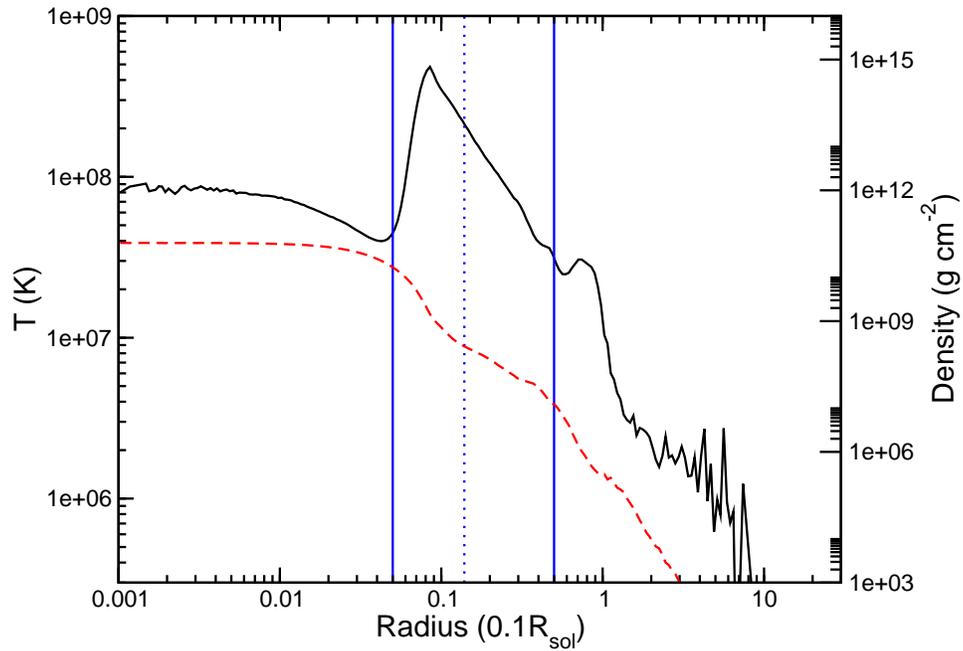}
  \caption{Radially averaged temperature (solid line) and density (dashed line) profiles of the final He+CO white dwarf merger product as a function of radius (in units of 0.1 solar radii). The temperature profile clearly shows three regions: a central object, the hot corona characterised by a temperature spike, and the accretion disk. The vertical lines indicate the limit of the convective mixing region (solid representing ``deep'' mixing and dotted for ``shallow'' mixing in the hot corona). {\it See the online edition of the journal for a colour version of this figure}.}
  \label{fig:RadProfile}
\end{figure}

A network containing 14 nuclei was employed in the SPH simulations to
account for the energy released by nuclear reactions during the merger
(mostly $\alpha$-capture reactions), while the detailed
nucleosynthesis reported here is calculated for tracer particles using
a 327 nucleus post-processing network ranging from hydrogen to
gallium.  {Energy generation rates between the two networks differs by
less than one percent for our models.}  The thermonuclear reaction
rates are adopted from the REACLIB database \citep{REACLIB} with
updates on experimental rates from \cite{ILI10a}.  SPH particles that
reside in the hot corona of the final object will characterise the
``atmosphere'' that is observed in a hydrogen deficient star.  The
tracer particles used to follow the thermodynamic history of the
merger are therefore picked from $0.005 < R/R_{\sun} < 0.05$ as shown
in figure \ref{fig:RadProfile}.  This range includes particles that
represent everything from the surface of the central dense object to
the inner edge of the accretion disk.  A total of 10\,000 tracer
particles have been used here to ensure a representative sample from
the two white dwarfs.

Our simulations take into account the existence of thin helium and
hydrogen buffer shells, which are predicted to survive prior evolution
through the common envelope stage. Therefore, three different regions
(carbon/oxygen-, helium-, and hydrogen-rich) are defined in the CO
white dwarf, while two (helium- and hydrogen-rich) are adopted for the
He white dwarf.  These regions are defined according to the mass
contained.  Adopting the notation of \cite{SAI02} --- who, for
instance, referred to the mass of the He shell in a CO white dwarf as
$M_{\textrm{He:CO}}$ --- the different shells are defined as follows:
$M_{\textrm{CO:CO}} = 0.78$, $M_{\textrm{He:CO}}= 0.019$,
$M_{\textrm{H:CO}} = 0.001$, $M_{\textrm{He:He}}= 0.399$,
$M_{\textrm{H:He}} = 0.001$.  The chemical composition of each shell
has been taken from detailed pre-white dwarf models \citep{REN10} of
the appropriate metallicity, with some additional information from
\cite{JEF11} for the hydrogen buffers.  It consists of 7 species:
$^{1}$H, $^{4}$He, $^{12}$C, $^{14}$N, $^{16}$O, $^{17}$O, and
$^{22}$Ne.  {It is worth noting that, although we do not expect to
reproduce the observed s-process abundances in hydrogen-deficient
stars with only these 7 isotopes, they should be enough to model
nucleosynthesis of light and intermediate-mass elements (up to neon).}
Two metalicity scenarios have been considered: (i) solar metalicity
and (ii) a lower metalicity one, with $Z=10^{-5}$.

To account for convective mixing of the material in the hot corona, we
calculate the mass-averaged abundances of all particles that reside
within a defined convective zone at the end of the hydrodynamic
simulation. Although the SPH calculations suggest that the entire hot
corona will be convective, we adjust the depth of the mixing in our
analysis to study the effect of this assumption on the final surface
abundances. Here, we consider two cases, (i) ``deep'' mixing, in which
everything in the range $0.005 < R/R_{\sun} < 0.05$ is mixed
homogeneously, and (ii) ``shallow'' mixing, in which the range is
assumed to be smaller, $0.014 < R/R_{\sun} < 0.05$.

\section{Results and Discussion}
\label{sec:nucleosynthesis}

\begin{figure} \centering
  \includegraphics[width=0.5\hsize]{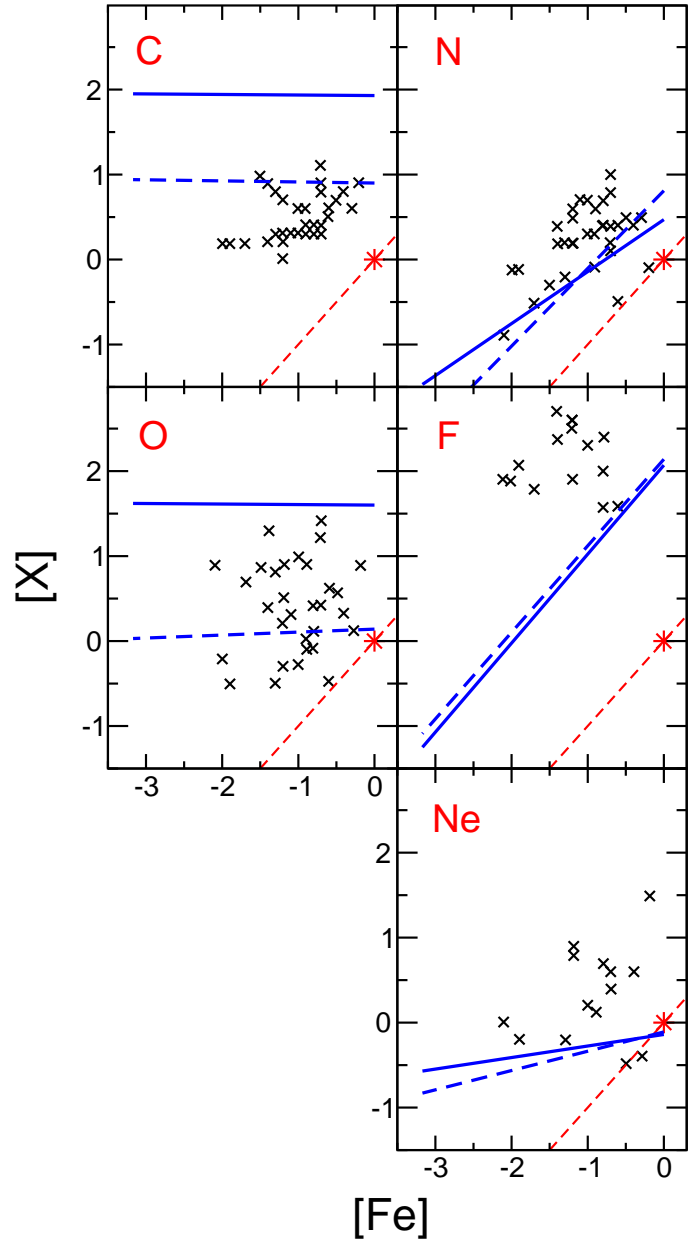}
  \caption{Hydrogen deficient star surface abundance determinations taken from \cite{JEF11} compared to our nucleosynthesis yields computed in the merger of a $0.4\, M_{\sun}$ and a $0.8\, M_{\sun}$ white dwarfs.  The axis labels, [X] and [Fe], correspond to the logarithmic abundances relative to solar for individual elements and for iron, respectively.  Solid lines are our calculations assuming ``deep'' convective mixing, while dashed lines are for ``shallow'' mixing (see text). The lines are added to guide the eye since our calculations include only two metallicities: $Z=0.015$ and $Z=10^{-5}$.  The asterisk symbol corresponds to solar abundances, while the diagonal lines intersecting solar values are the abundances expected if the solar values were scaled with metalicity.  {\it See the online edition of the journal for a colour version of this figure}. }
  \label{fig:results}
\end{figure}

The mass averaged abundances obtained from the tracer particle
nucleosynthesis are compared with the observed abundances in figure
\ref{fig:results}.  The plotted solid lines correspond to the final
abundances obtained under the ``deep'' mixing assumption, while
surface abundances obtained assuming a ``shallow'' mixing depth are
represented by dashed lines. These lines are added purely to guide the
eye since our calculations were only performed for two metalicity
cases ($z=0.015$ and $z=10^{-5}$).  In judging the successfulness of
our models, it is worthwhile to compare each element to observed
abundances on a one by one basis.

Carbon depends strongly on the mixing depth assumed in the hot corona
of the merger product but appears to be consistently high in our
calculations. Nevertheless, the overall trend of carbon abundance with
respect to metalicity agrees with the observational data.  {The
$^{13}$C$/^{12}$C ratio for solar metalicity is $\sim 2 \times
10^{-5}$}, consistent with the predicted nature of helium-burned
material, but considerably higher than the observed ratio.  This
disagreement could arise from the limited number of nuclear species
adopted in our calculations.

The oxygen abundance also agrees fairly well with the observed data,
showing a large range of values depending on the mixing depth
assumed. This dependence could explain the observed scatter in oxygen
in EHe and RCB stars, where the observed oxygen abundance could depend
strongly on the nature of the individual merging event. This picture
is not, however, consistent with the relatively low scatter in carbon
abundances.  With regard to the isotopic abundances, our solar
metalicity model for deep and shallow mixing yields
$^{16}$O/$^{18}$O~$= 370$ and $19$, respectively.  The $^{18}$O has
clearly been enhanced in the outer regions of the hot corona with
respect to solar abundances (where $^{16}$O/$^{18}$O~$\approx 400$),
although not to the extent found in some HdC and RCB stars
\citep{CLA07}.

Nitrogen follows the expected trend indicating that it is enhanced
only through CNO cycling in the parent stars. Although it is slightly
reduced by the \reaction{14}{N}{$\alpha$}{$\gamma$}{18}{F} reaction,
it is present in such a high concentration that it is barely
diminished in the short time-scales involved.

The overabundances of fluorine obtained in our models are high, but
not high enough to fully agree with the observational data. The lack
of sensitivity to mixing depth is an indication that fluorine is
produced homogeneously in the merging event, and not in any particular
region of the merger. Fluorine is produced dominantly through the
following sequence of reactions:
$^{14}$N($\alpha$,$\gamma$)$^{18}$F(p,$\alpha$)$^{15}$O($\alpha$,$\gamma$)$^{19}$Ne($\beta^+$)$^{19}$F.
Enrichment of fluorine in our models is a great success, suggesting
that hydrogen deficient stars could be the result of a \textit{hot}
white dwarf merger event since it is challenging to reconcile those
high abundances with a cold merger event.

Finally, the abundance of neon following the merger is consistent with
the metallicity, indicating that it is not processed in either the
merger event, or during the preceding evolution of the parent
stars. Neither our models, or the cold merger models of \cite{JEF11}
reproduce the observed neon abundance in several EHe stars, which
contain overabundances up to about 1 dex. Clearly more work must be
performed to understand the nucleosynthesis of neon in hydrogen
deficient stars.

In summary, our models of hot white dwarf mergers are able to
reproduce, on an order of magnitude level, the surface abundances of
light and intermediate-mass nuclei in hydrogen deficient stars. The
model's main success is in its ability to synthesise high fluorine
abundances, which cannot be produced in cold merger models
\citep{JEF11}.

\section{Conclusions}
\label{conclusions}

Mergers of helium and carbon-oxygen white dwarfs are currently favoured
as the events responsible for the production of hydrogen deficient
HdC, EHe, and RCB stars. The main question facing us is whether these
objects are the result of a cold merger, in which no nucleosynthesis
occurs, or a hot merger, in which at least some of the material is
processed through nucleosynthesis.

The results presented in this work show that nuclear processing during
the merger event can account for some of the observed abundances in
these stars, particularly for fluorine, which cannot be easily
synthesised in the cold merger scenario.  There are, however, still
open questions regarding the expected nucleosynthesis: neon is
difficult to produce at the observed levels, while the absolute
abundance of carbon is too high in our models. Such discrepancies may
be alleviated by the inclusion of a more detailed initial chemical
composition with many more species, resulting from state-of-the-art
models of the preceding AGB phase.  Inclusion of such detailed
abundances is currently under way.

\acknowledgements

This work has been partially supported by the Spanish grants
AYA2010-15685, AYA08-1839/ESP, and AYA2008- 04211-C02-01, by the
E.U. FEDER funds, and by the ESF EUROCORES Program EuroGENESIS through
the MICINN {grants EUI2009-04167 and 04170}.\\

\bibliographystyle{aa}

\end{document}